# Potential for ultrafast dynamic chemical imaging with few-cycle infrared lasers

**Toru Morishita[1,2], Anh-Thu Le[1], Zhangjin Chen[1] and C D Lin[1]**

[1]Department of Physics, J. R. Macdonald Laboratory, Kansas State University, Manhattan, KS, 66506, USA
[2]Department of Applied Physics and Chemistry, University of Electro-Communications, 1-5-1 Chofu-ga-oka, Chofu-she, Tokyo, 182-8585, Japan

E-mail:cdlin@phys.ksu.edu, toru@pc.uec.ac.jp, atle@phys.ksu.edu

**Abstract.** We studied the photoelectron spectra generated by an intense few-cycle infrared laser pulse. By focusing on the angular distributions of the *back* rescattered high energy photoelectrons, we show that accurate differential elastic scattering cross sections of the target ion by *free* electrons can be extracted. Since the incident direction and the energy of the *free* electrons can be easily changed by manipulating the laser's polarization, intensity, and wavelength, these extracted elastic scattering cross sections, in combination with more advanced inversion algorithms, may be used to reconstruct the effective single-scattering potential of the molecule, thus opening up the possibility of using few-cycle infrared lasers as powerful table-top tools for imaging chemical and biological transformations, with the desired unprecedented temporal and spatial resolutions.

## 1. Introduction

When an object is illuminated by a plane wave, the amplitude of the diffraction pattern in the far field is the Fourier transform of the object. To achieve high resolution, microscopists use short wavelengths, such as X-rays or high-energy electrons, with energies of tens to hundreds of keV, to obtain diffraction patterns of the object. For image reconstruction, advanced phase retrieval algorithms have been developed [1,2].

While these are powerful tools for spatial imaging of structure at the atomic levels, they are not adequate for studying chemical and biological transformations that involve complex transient structures [3,4]. Since these reactions occur within a few to hundreds of femtoseconds, techniques must be developed to observe the coherent atomic motions in the corresponding time scale. As existing X-ray diffractions using light sources from synchrotron radiation have pulse durations of tens of picoseconds, they are not suitable for time resolved imaging. The impending completion of the next generation X-ray free-electron lasers (XFELs) promises to offer extremely short bursts of X-rays with high intensity and duration of several femtoseconds. While these facilities may offer the opportunity to perform time-resolved X-ray diffraction of chemical and biological reactions in the future, they are large and costly facilities. Other ideas and possible new techniques for imaging transient molecules remain highly desirable.

The advent of femtosecond (fs) lasers in the past decades has made it possible to study and control chemical and biochemical reactions. Today laser pulses of duration of a few femtoseconds are easily accessible. However, these infrared lasers have wavelength much longer than the interatomic distances, hence they are considered not suitable for imaging ultrafast structural changes in molecules. To overcome this limitation, ultrafast electron diffraction (UED) method has been developed [3,4] where electron pulses are generated from femtosecond lasers and then accelerated to about 120 keV. Most recently, UED with temporal resolution of about 100 fs has been used to observe the time-dependent structural change in phase transition [4].

In this article we present a different approach for achieving ultrafast laser induced electron microscopy (LIEM) with infrared lasers. Here the temporal resolution is controlled by the pulse duration of the laser, while the structural information of the target is to be retrieved from laser-induced photoelectron spectra. For the latter to work, two ingredients are needed. First, the electron released from the molecule by the laser has to return to recollide with the target ion. From these recollisions elastic differential scattering cross sections of the target ion by *free* electrons are extracted. An infrared laser can be easily manipulated to change its polarization direction, intensity and wavelength, thus a large set of elastic scattering cross sections are readily collected. Second, a new inversion algorithms should be developed to reconstruct the single-scattering potential from the elastic scattering cross sections, from which the structural information on the molecule is derived. Since the returning electrons have energies of several tens to hundreds eV's only, the elastic scattering cross sections are not given by the Fourier transform of the object as in the conventional diffraction theory.

The possibility of using infrared lasers for imaging molecules has been discussed in the literature [5], but later theoretical investigations have revealed many difficulties [6-9]. Consider a typical five-cycle laser pulse, with mean wavelength of 800 nm and peak intensity at $10^{14}$ W/cm$^2$. The electric field $\vec{E}(t) = -\partial \vec{A} / \partial t$ and the vector potential $\vec{A}(t)$ of such a pulse are depicted in Fig. 1a. At time near "a" where the electric field strength is large, an electron is tunnel ionized. Schematically, this electron is first accelerated to the right ( Fig.1b), but it may be driven back in the next half cycle to recollide with the target ion, at time near "b". This revisit incurs various electron-ion collision phenomena, the dominant one being elastic scattering by the target ion. This rescattering forms the basis of self-diffraction proposed by Corkum and co-workers [10]. Unlike previous works where the electron images were analyzed in the forward directions, here we demonstrate that the route to the success of LIEM requires experimentalists to take momentum images of electrons that are rescattered into the backward directions. At the time of recollision near "b", the returning electrons have energies of about 3.17 $U_P$. Here $U_P$ is the ponderomotive energy, i.e., the mean quiver energy of an electron in a laser field. For electrons that are rescattered into the forward directions (the left-side of Fig. 1b), they will be decelerated by the laser field in the next half cycle and emerge at low energies. Unfortunately, low-energy electrons on this side are also generated by tunnelling ionization near "a' " , about half a cycle after "a". These electrons would have energies up to about 2 $U_P$, and would interfere with the forward rescattered electrons to damage the diffraction images. On the other hand, electrons that are rescattered into the backward directions will be accelerated by the laser field in the next half cycle, resulting in much higher energies of up to about 10 $U_P$[11]. These high-energy electrons, which have been observed previously [12], would preserve the diffraction images and can be used to retrieve the structure of the target ion as probed by the rescattering electrons.

## 2. Extracting accurate electron scattering cross sections from laser-induced photoelectron momentum spectra

Accurate high-energy photoelectron momentum spectra of molecules by infrared lasers are not available either experimentally or theoretically so far. For atomic targets, these spectra can be accurately calculated by solving the time-dependent Schrödinger equation. In Fig. 2a we show the electron momentum images of the simplest atomic hydrogen target generated by a typical infrared laser, chosen for a five-cycle pulse with peak intensity of $5x10^{13}$ W/cm$^2$ and mean wavelength of 800nm. Since the electron yield drops very rapidly with energy [Fig. 2(c)] the momentum images in Fig. 2a for each photoelectron energy have been renormalized, i.e., the yield integrated along each circle of the $p_z$ -$p_x$ plane is unity ($p_z$ is along the direction of polarization). We pay attention only to the momentum distributions of high energy electrons. The images show two clear half circular rings, or ridges, one on the "left" side with smaller radius, and another on the "right" with larger radius, where the center of each half circle is shifted from the origin of the $p_z$ -$p_x$ plane. We called these circular rings back rescattered ridges (BRR), representing electrons that have been rescattered into the backward directions by the target ion. The BRR on the "right" is from electrons born at time near "a" (Fig.1a), travelling to the "right" by the laser field (Fig.1b) and then returning to the target ion at time near "b", where they are rescattered back to the right. Each momentum half circle is represented approximately by $A_r \hat{p}_z + p_o \hat{p}_r$, where the second term represents the momentum vector of the elastically rescattered electron with kinetic energy given by $3.17 U_P = p_0^2 / 2$, while the first term represents the momentum shift due to the drift of the electron from "b" where the vector potential is $A_r$ to the end of the laser pulse. (We used atomic units here where the vector potential has the units of momentum. More precise fitting shows that the shift is 0.95 $A_r$. ) Similarly, the BRR on the left is due to electrons that were born near "a' " and rescattered into the backward directions near "b' " (Fig.1a). Due to the smaller $A_r$ at "b' ", the shift of the center and the radius of the circle are smaller.

Fig. 2b shows the same normalized momentum images of Xe using identical laser parameters. Here Xe is treated in the single active electron model. Similar shifted half circular rings are seen, but along each BRR, the intensity exhibits a clear dip away from the polarization axis. Can one attribute the difference in Xe and H targets to rescattering alone, i.e., to the difference in the elastic scattering cross sections by free electrons only? Taking the *actual* calculated electron yields on the BRR, i.e., along $A_r \hat{p}_z + p_o \hat{p}_r$, we compare the angular dependence of the momentum images on the BRR with the elastic differential cross sections of $Xe^+$ with "free" electrons ( Fig. 2f) at incident energy $E_0 = p_0^2 / 2$, with the scattering angle $\theta$ measured from the shifted center, The two curves are in excellent agreement. Similar agreement is shown at the same peak intensity for atomic hydrogen in Fig. 2d and for Ar in Fig. 2e. We have also tested other atoms and other intensities and found similar agreement.

In calculating the electron momentum spectra by intense laser pulses, we solved the time-dependent Schrödinger equation (TDSE) of a one-electron model atom in the laser field [13,14]. Care was taken to make sure that the momentum spectra near 10 $U_P$ and beyond are converged. A sine-squared pulse with carrier-envelope phase equal to zero is assumed for the laser pulses. The same model potential used for solving the TDSE is also used to calculate the elastic scattering cross sections by free electrons. Since the atomic ions are assumed to be structureless, the differential elastic scattering cross sections are obtained from the simple potential scattering theory by calculating partial wave phase shifts.

We comment that the results shown in Fig. 2 are from calculations based on a five-cycle pulse with carrier envelope phase (cep) $\phi = 0$. For short pulses, it is essential that the cep be stabilized for the duration that the electron spectra are taken [15]. For a different $\phi$, the value of the vector potential at the return time is different. For long pulses, electrons born at successive cycles contribute to the momentum spectra, and the yield along the BRR will exhibit interference typical of above-threshold-ionization electrons. In this case, the elastic scattering cross sections can be extracted from the envelope of the electron yield along the BRR. In other words, elastic scattering cross sections of atomic ions by free electrons can be extracted from rescattering generated photoelectron spectra on the BRR, for long or short laser pulses, and for different intensities and wavelengths.

## 3. Implications for molecular targets

Based on the above results which were obtained from "exact" theoretical calculations, we conclude that laser induced photoelectron momentum images on the BRR allow us to extract elastic differential scattering cross sections of the target ion by *free* electrons. While calculations have been carried out only for atoms, there is no reason to believe that the conclusion would not hold for molecules. At present, "exact" TDSE calculations for the momentum images along the BRR for molecules is not available. Thus we used the second-order strong field approximation (SFA2) for molecular targets. The momentum image along the BRR calculated using SFA2 for atomic targets have been shown [16] to be in good agreement with the elastic scattering cross sections calculated using the first Born approximation. This has been shown to be true for molecular targets as well. In Fig. 3, on the top frame we show the differential elastic scattering cross sections of free electrons incident on $H_2^+$ with its internuclear axis lying perpendicular to the direction of the incident beam. The momentum of the incident electron is 2.2 a.u., and the internuclear separation of $H_2^+$ is R=3 a.u. Taking the incident direction as the z-axis, the differential cross sections are expressed in terms of spherical angles $\theta$ and $\phi$. In the bottom frame, we show the calculated photoelectron momentum spectra for $H_2^+$ in the same geometry but the electron is being replaced by a 5 fs, 800 nm, laser pulse with peak intensity of $3 \times 10^{14}$ W/cm$^2$, with the laser polarized along the z-axis. The extracted photoelectron momentum distributions along the BRR clearly are in good agreement with the differential elastic scattering cross sections of the same target by free electrons. We expect similar agreement when "exact" calculations are carried out, as in atoms, or when experimental data become available.

For molecules in general, this conclusion is far reaching since it would allow experimentalists to "measure" elastic scattering cross sections by electrons using infrared lasers. Since electron scattering is a powerful tool for determining the structure of atoms and molecules in the traditional energy-domain measurements, this implies that infrared lasers can be employed to probe the detailed structure of molecules. For few-cycle infrared lasers, the rescattering of the electrons with the parent ions occurs within the order of one femtosecond and with attosecond temporal resolution. Thus few-cycle infrared lasers can handily achieve subfemtosecond temporal resolution, and they may serve as efficient ultrafast cameras for imaging physical, chemical and biological systems where the system is undergoing rapid structural change.

**4. Spatial resolution achievable with infrared lasers**

Given that elastic scattering cross sections can be extracted from laser-induced electron momentum spectra, the next question is whether molecular structure can be retrieved with good resolution from these data in view that the energies of the rescattering electrons (of the order of 100 eV or less) are much smaller than those used in conventional electron diffraction experiments (of the order of 10-100 keV). We think it is possible since in the present scheme electron images are taken only in the backward directions. To make such large-angle scattering the electrons have to be scattered off from each atomic center in the molecule. The scattered waves from these centers generate diffraction images which will be analyzed to retrieve the positions of the atoms.

Before a retrieval program becomes available, we first show that the momentum images of these backscattered electrons indeed are quite sensitive to the arrangement of atoms in a molecule. In Fig.4 elastic differential cross sections of $H_2^+$ at different internuclear distances are shown. The electron momentum is p=2.2 a.u. and the internuclear distances are $R$=2, 3, 4, and 5 a.u. respectively. The incident electron beam is perpendicular to the internuclear axis. One can clearly see that the positions of the interference minima/maxima shift as $R$ increases. Also, as $R$ increases, more interference minima/maxima emerge. In fact, within the first Born approximation, the differential cross section can be written as

$$\sigma(\theta, \varphi) \sim \cos^2[pR\sin(\theta)\cos(\varphi)].$$

Therefore, destructive interference from the two atomic centers occurs when

$$pR\sin(\theta)\cos(\varphi) = (\frac{1}{2} + n)\pi,$$

$$n = 0, \pm 1, \pm 2 ...$$

The positions of the interference minima shown in Fig. 4 are in very good agreement with the above equation. Note that the diffraction pattern changes quite rapidly with respect to R , thus we anticipate that it would be relatively easy to achieve sub-Angstrom spatial resolution when the internuclear distance is retrieved from these data.

As another example, in Fig. 5 we show the differential elastic scattering cross sections from two planar molecules, benzene and nitrobenzene, respectively, with the incident electron direction perpendicular to the molecular plane and with momentum at p=2.5 a.u. From the momentum images, those that are mostly due to the benzene backbone can be identified. Additional features due to the extra nitrogen atoms can also be clearly seen. These diffraction images are taken at the relatively low electron energies and may be used to retrieve the structural information.

For dynamic chemical imaging, in Fig. 6 we show how the elastic scattering cross sections vary as the $C_2H_2$ molecule goes through intermediate steps in the isomerization process--- starting from vinylidene and passing through two transition states TS1 and TS2 and a local minimum M1, and ending up at the global minimum state, acetylene. This structural rearrangement is depicted along the left column. The momentum images show some differences, with the most noticeable difference in vinylidene as compared to acetylene. The images were taken for incident electrons perpendicular to the molecular plane. In this example, the images for the intermediate states do not differ significantly since only the light hydrogen atoms undergo position changes mostly. It would be a critical test of the retrieval method in the future to check if such small changes in the images can still be retrieved to obtain accurate molecular structure.

In the above examples, electron momentum images were taken from incident electrons coming from one side only. Consider a moderate-size asymmetric molecule fixed in space. If infrared lasers are used to illuminate this molecule from three perpendicular directions, by collecting momentum images in the backward rescattering directions along each axis, many sets of elastic differential cross sections can be extracted. Additional data can be obtained from different laser intensities or wavelengths, including lasers with wavelength up to about 2 microns [17]. Such a wealth of elastic differential scattering cross sections would then be available for retrieving the structure of the molecule.

## 5. Promises and challenges

In this paper we have shown that high-energy electron momentum spectra along the back rescattering ridge (BRR) induced by few-cycle laser pulses can be used to extract accurate differential elastic scattering cross sections of the molecular ions by the electrons generated by the laser itself. These differential cross sections can in principle be used to retrieve the structure of the molecules. Since the pulse duration of few-cycle infrared pulses is in the order of a few femtoseconds and rescattering occurs within the attosecond scale, we illustrated that such few-cycle infrared laser pulses can be used for dynamic chemical imaging of transient molecules.

Experimentally the momentum spectra on the BRR can be measured with COLTRIMS detectors [18] where momentum images over $4\pi$ angles can be collected, or by using the time-of-flight methods. Since the yields of high-energy electrons on the BRR are small, one may want to enhance counting rates by increasing gas pressure, by increasing the repetition rates of the laser pulses, or by increasing laser intensities. In the future, one may want to look into other optical methods, such as superposing attosecond pulse trains [19], to enhance the yield of the returning electrons and to increase the electron counts on the BRR.

The calculations presented so far are for molecules fixed in space. For molecules in the gas phase, they are isotropically distributed initially. To retrieve structural information, they should be at least partially aligned before electron spectra are taken. Molecules can be aligned in 1D with one laser or in 2D or 3D with two or more lasers [20,21]. The degree of alignment of small molecules can be measured by Coulomb explosion independently, or calculated theoretically [20]. If the molecules are isotropically distributed then partial information, such as the internuclear separation may still be extracted (but not the bond angles), as often been done from electron diffraction experiments [22].

## Acknowledgments

This work was supported in part by the Chemical Sciences, Geosciences and Biosciences Division, Office of Basic Energy Sciences, Office of Science, U. S. Department of Energy. TM is also supported by a Grant-in-Aid for Scientific Research (C) from the Ministry of Education, Culture, Sports, Science and Technology, Japan, by the 21st Century COE program on ``Coherent Optical Science'', and by a Japanese Society for the Promotion of Science (JSPS) Bilateral joint program.

**Figure legends**

**Figure 1**.   (a) The electric field (E) and the vector potential (A) of a typical five-cycle infrared laser pulse.  (b) Schematic of using backward rescattering process to image the molecule by its own electron.

**Figure 2**. Electron momentum spectra generated by a  5-cycle laser pulse, with peak intensity of $5 \times 10^{13}$ W/cm$^2$ and mean wavelength of 800 nm on the  $p_z$-$p_x$ plane for (a) H and (b) Xe. The images are renormalized for each photoelectron energy to reveal the global angular distributions.  The electron energy distributions for H are shown in (c). The electrons along the backward rescattering ridges (BRR) are identified by the outermost circular rings on each side. Comparison of the angular distributions of electron images along the BRR with the differential elastic scattering cross sections by free electrons, (d) H, (e) Ar and (f) Xe, all under the same laser intensity, illustrating that the latter can be extracted from the laser-generated electrons along the BRR.

**Figure 3.** Comparison of (a) elastic differential scattering cross sections of $H_2^+$ by electrons with (b) the same information obtained from laser-induced electron images along the BRR, showing good agreement between the two. The molecular axis is perpendicular to the direction of the incident electron, so is the direction of the laser polarization. Electron momentum is *p=2.2* a.u.,  corresponding to the BRR obtained with a four-cycle laser with the peak intensity of $3 \times 10^{14}$ W/cm$^2$, wavelength of 800nm.

**Figure 4.** Elastic scattering cross sections as in Fig.3, but at four different internuclear distances.

**Figure 5.**  Differential elastic cross section for electron scattering from benzene (top panel) and nitrobenzene (bottom panel), at the collision momentum of *p=2.5* a.u., as a function of scattering angles θ and φ. The incident beam is perpendicular to the molecular plane in both cases.

**Figure 6.**  Schematic of dynamical chemical imaging of the isomerization of acetylene/vinidyne using infrared lasers. The isomerization of the linear acetylene molecule (bottom) to the planar vinylidene (top) is assumed to pass through three intermediate/transition states, with their geometries depicted along the left column. For these geometries, the differential elastic scattering cross sections extracted from the BRR (*p=2.2* a.u.) are shown for laser pulses

polarized perpendicular to the molecular plane. These images can be used to retrieve the time-dependent structural rearrangement of atoms in the isomerization process.

(a)

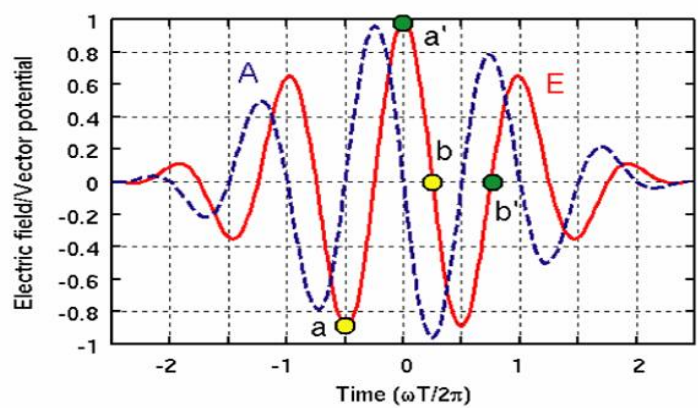

(b)

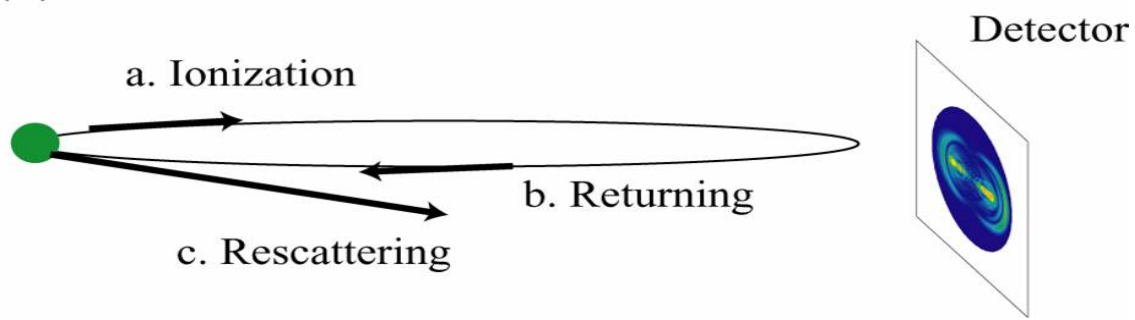

Figure 1

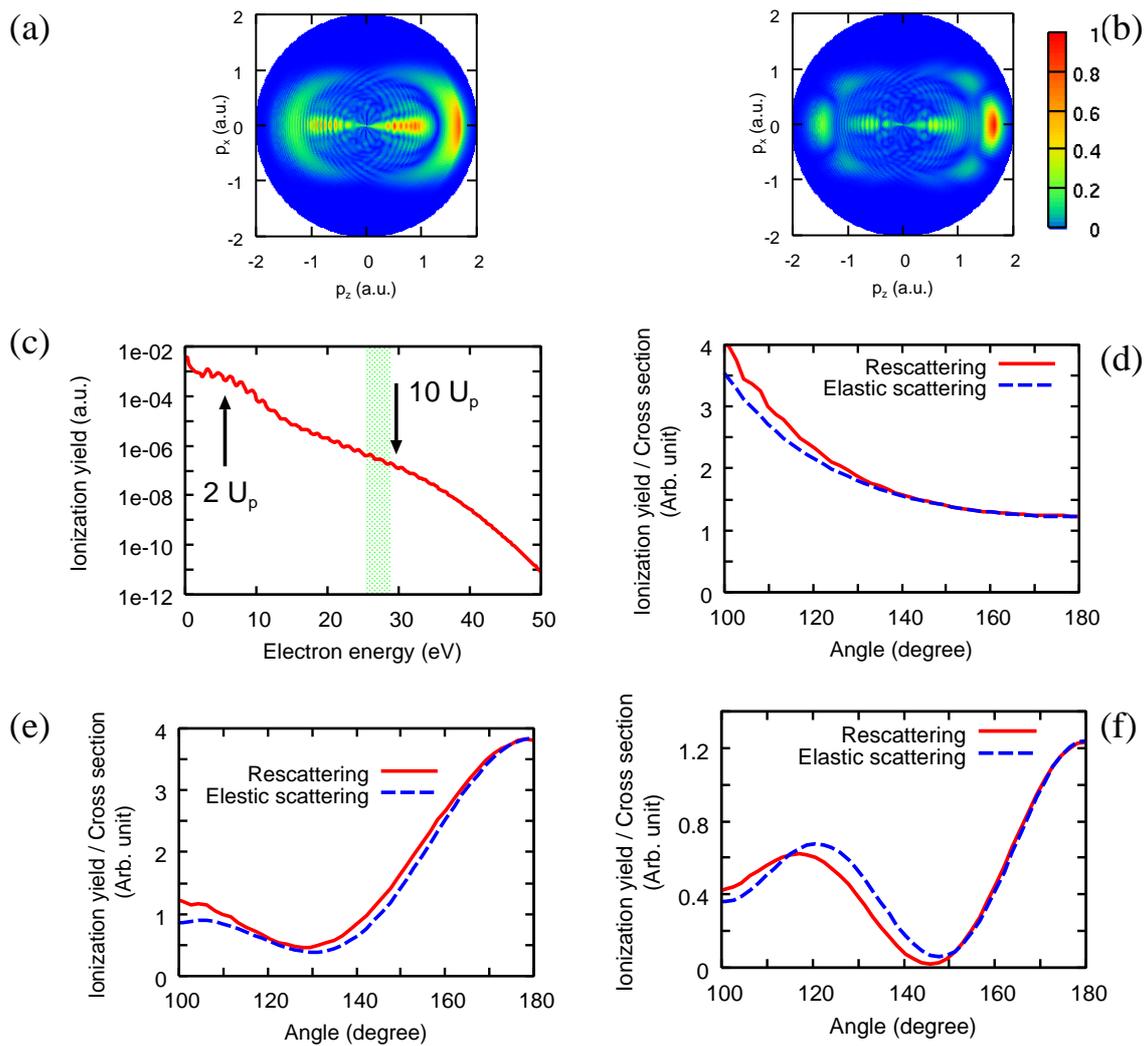

Figure 2

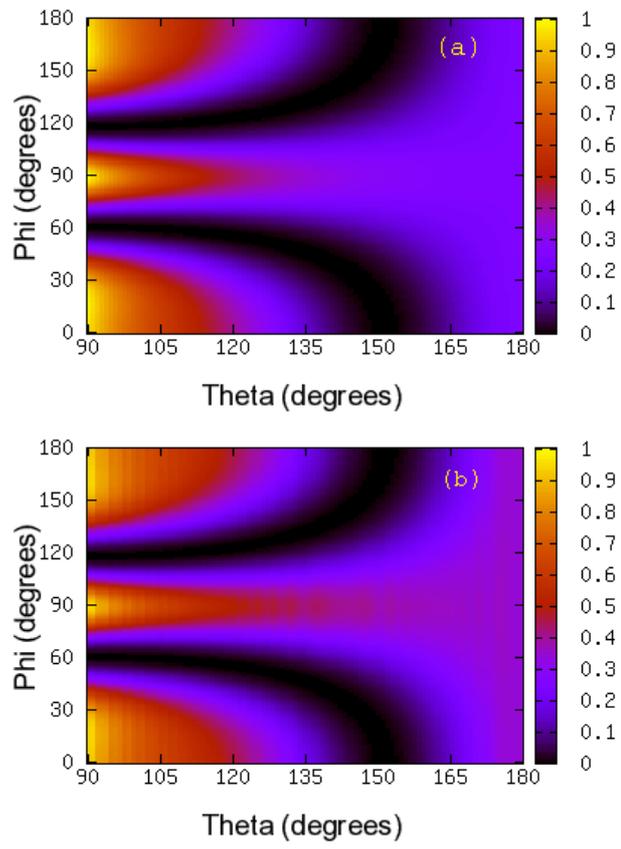

Figure 3.

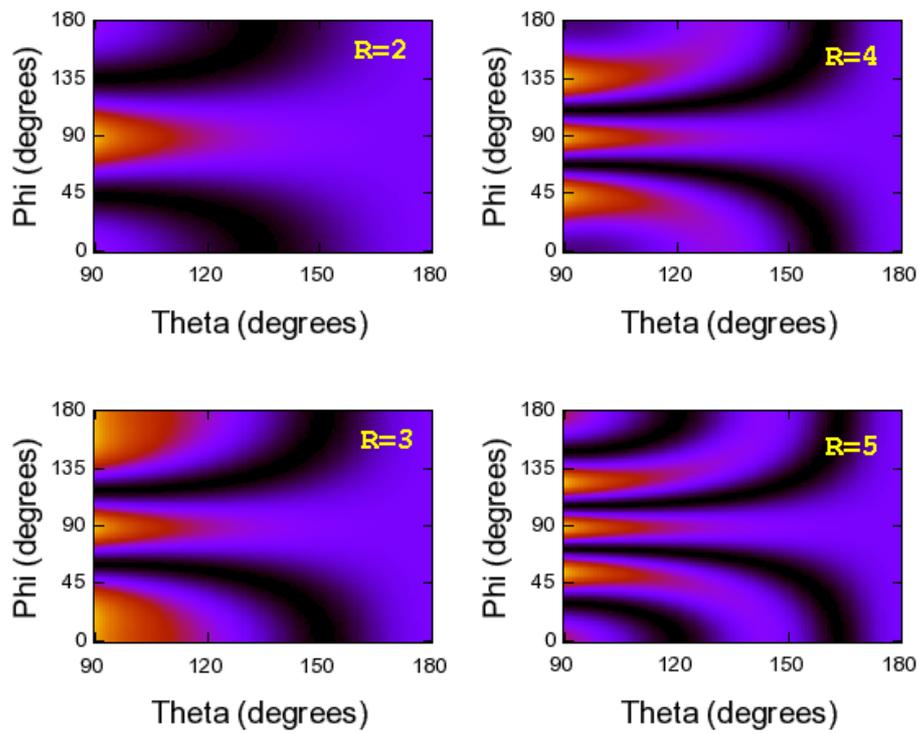

Figure 4.

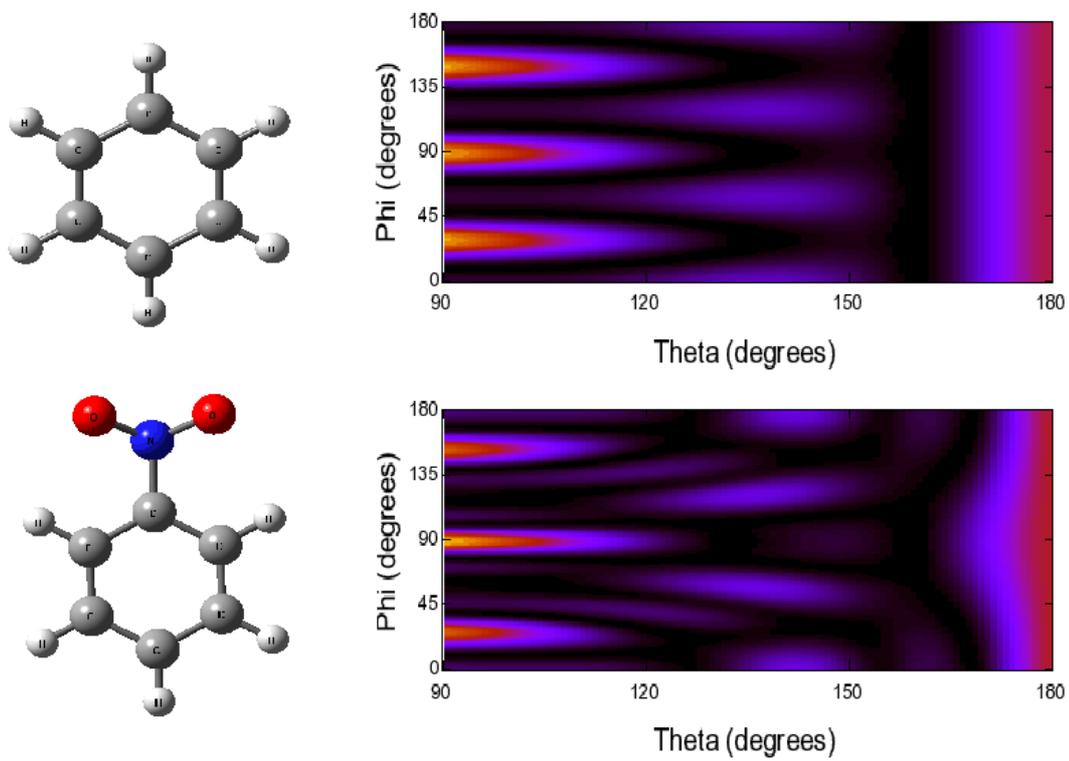

Figure 5.

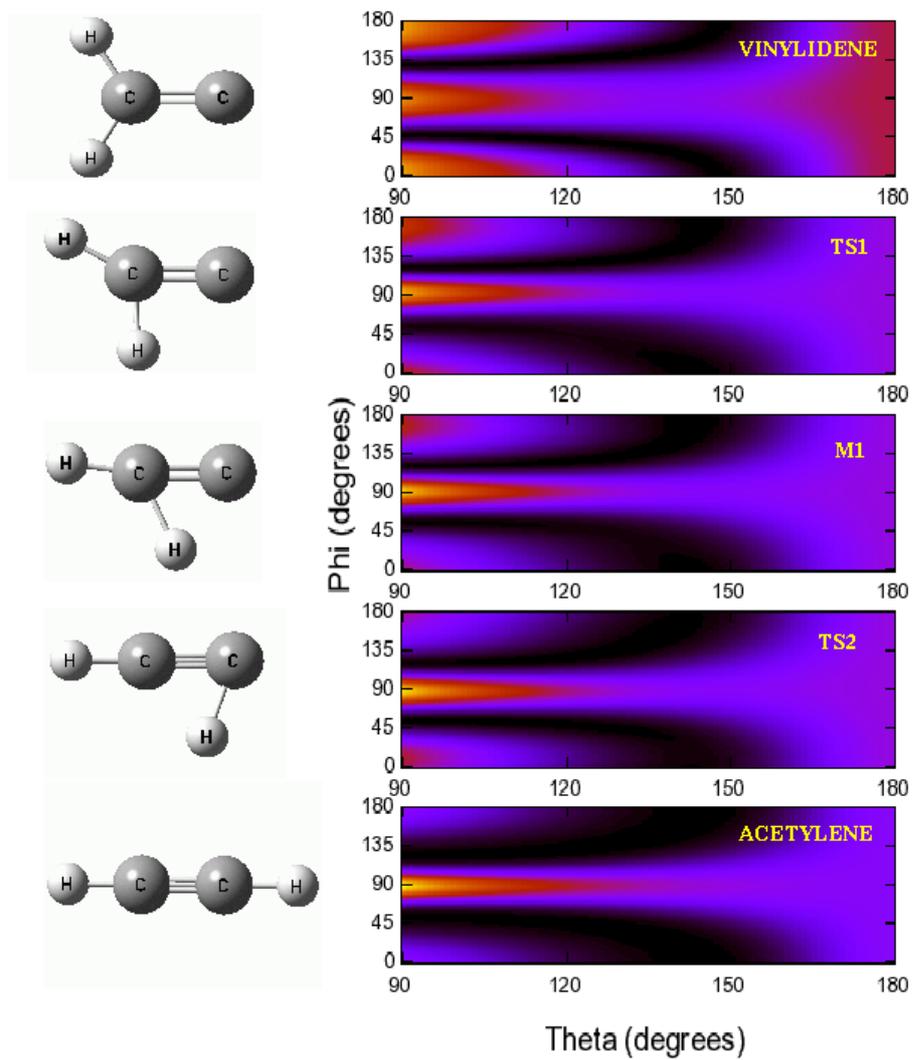

Figure 6.